\documentclass[twocolumn,floatfix,showpacs,prl,superscriptaddress]{revtex4-1}
\usepackage{graphicx}
\usepackage{amsmath}
\usepackage{amssymb}
\usepackage{bm}
\usepackage{hyperref}
\usepackage{multirow}
\usepackage{color}

\usepackage{etoolbox}

\newcommand{\pd}{{\phantom\dag}}

\begin{document}

\title{Statistical translation invariance protects a topological insulator from interactions}
\author{A. Milsted}
\affiliation{Institut  f\"{u}r Theoretische Physik, Leibniz Universitat Hannover, Appelstrasse 2, 30167 Hannover, Germany}
\author{L. Seabra}
\affiliation{Department of Physics, Technion -- Israel Institute of Technology, Haifa 32000, Israel}
\author{I. C. Fulga}
\affiliation{Department of Condensed Matter Physics, Weizmann Institute of Science, Rehovot 76100, Israel}
\author{C. W. J. Beenakker}
\affiliation{Instituut-Lorentz, Universiteit Leiden, P.O. Box 9506, 2300 RA Leiden, The Netherlands}
\author{E. Cobanera}
\affiliation{Institute for Theoretical Physics, Center for Extreme Matter and Emergent Phenomena,
Utrecht University, Leuvenlaan 4, 3584 CE Utrecht, The Netherlands}
\date{April 2015}
\begin{abstract}
We investigate the effect of interactions on the stability of a disordered, two-dimensional topological insulator realized as an array of nanowires or chains of magnetic atoms on a superconducting substrate.
The Majorana zero-energy modes present at the ends of the wires overlap, forming a dispersive edge mode with thermal conductance determined by the central charge $c$ of the low-energy effective field theory of the edge.
We show numerically that, in the presence of disorder, the $c=1/2$ Majorana edge mode remains delocalized up to extremely strong attractive interactions, while repulsive interactions drive a transition to a $c=3/2$ edge phase localized by disorder.
The absence of localization for strong attractive interactions is explained by a self-duality symmetry of the statistical ensemble of disorder configurations and of the edge interactions, originating from translation invariance on the length scale of the underlying mesoscopic array.
\end{abstract}
\pacs{03.65.Vf, 71.20.-b, 73.20.Fz}
\maketitle

\emph{Introduction} --- 
In the very first topological insulator ever discovered, the quantum Hall insulator \cite{Klitzing1980}, interactions have a dramatic effect by changing the quantum of conductance from the value $e^2/h$ to a fraction of it \cite{Stormer1999}.
We know now that the quantum Hall insulator is but one entry in a periodic table of topological states of matter, including both insulators and superconductors (and commonly referred to as topological insulators or TIs) \cite{Kitaev2009, Schnyder2009,Hasan2010,Qi2011}. 
These materials are all characterized by a gapped bulk with gapless surface or edge excitations, protected from localization by disorder due the existence of a topological invariant. 

Because a topological invariant is a property of the single-particle Hamiltonian, it is a challenge to classify TIs in the presence of interactions \cite{Liu2012, Chen2012, Chen2013, Heck2014, Senthil2014}. 
Besides the question of whether interactions will localize the surface states, one can also ask whether interactions may lead to distinct topological phases --- as they do in the fractional quantum Hall effect.
Here we address these two questions for the superconducting counterpart of the quantum Hall insulator, a two-dimensional (2D) superconductor with chiral \textit{p}-wave pairing.

The superconducting analogue of the electronic quantum Hall effect, the thermal quantum Hall effect, refers to a heat current carried by Majorana modes propagating along the edge of a topological superconductor \cite{Senthil2000, Read2000, Vishwanath2001}.
In the absence of backscattering along the edge, the thermal conductance equals $G_{\rm thermal}=c\,G_0$, with $G_0=\pi^2 k_{\rm B}^2T/3h$ the thermal conductance quantum for free electrons.
The coefficient $c$, called the central charge, governs the stress-energy tensor of the conformal field theory associated with the low-energy edge modes \cite{Blote1986, Affleck1986, Read2000, Sumiyoshi2013, Bardlyn2015}. 
While $c=1$ for both the integer and fractional quantum Hall effects, a Majorana edge has $c=1/2$ --- at least in the absence of interactions \cite{Read2000}.
One of our findings is that moderate repulsive interactions between the Majorana fermions drive a transition to an extended $c=3/2$ edge phase.

\begin{figure}[tb]
\includegraphics[width=0.9\columnwidth]{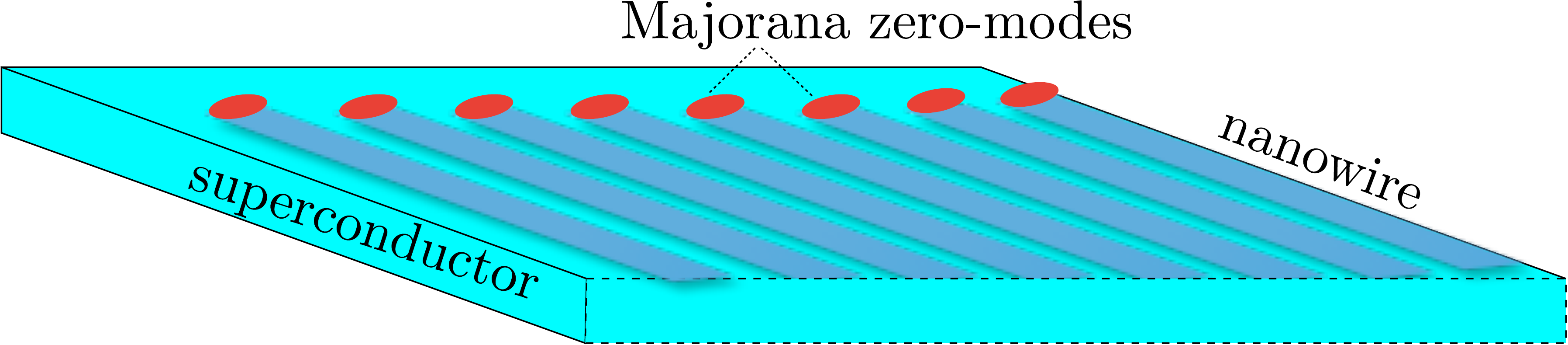}
\caption{Array of nanowires on a superconducting substrate, with a delocalized Majorana edge mode composed out of coupled zero-modes localized at the end points.
\label{fig:kitaevedge}}
\end{figure}

Chiral \textit{p}-wave superconductors may exist naturally (${\rm Sr}_2{\rm RuO}_4$ is a candidate material \cite{Mackenzie2003, Kallin2012}), but a major recent development is the search for this exotic pairing in semiconductor nanowires \cite{Mourik2012, Deng2012, Das2012, Rokhinson2012} and chains of magnetic atoms \cite{Nadj-Perge2014} deposited on a superconductor with conventional \textit{s}-wave pairing.
(Alternative proposals include chains of atoms in optical lattices \cite{Buhler2014} and chains of magnetic vortices on the surface of a 3D TI \cite{Chiu2015}).
A two-dimensional (2D) array of parallel nanowires, see Fig.~\ref{fig:kitaevedge}, forms an anisotropic topological insulator, called a ``weak'' TI \cite{Nomura2008, Ringel2012, Mong2012, Fu2012, Baireuther2014, Tanaka2012, Dziawa2012, Xu2012} because the Majorana mode propagates only along the edges perpendicular to the wires.
Each nanowire realizes a Kitaev chain \cite{Kitaev2001}, with two unpaired Majorana zero-energy modes at the end points of the wire.
These zero-modes overlap to form a dispersive 1D edge mode of Majorana fermions \cite{Wang2014,Wakatsuki2014,Seroussi2014}. Backscattering by disorder is not forbidden, yet this ``Kitaev edge mode'' does not localize \cite{Fulga2014,Diez2014a}.   

Because the effective boundary theory of the Kitaev edge is one-dimensional, it is possible to investigate its behavior in great detail with modern numerical tools based on matrix product state (MPS) methods (we use both variational infinite MPS methods and the density-matrix renormalization group --- DMRG).
The average translation symmetry of the mesoscopic array causes a statistical translation symmetry of the ensemble of disorder configurations and the exact self-dual structure of the interacting edge Hamiltonian.
This situation is an example of a ``statistical topological insulator", protected by a symmetry that is present only on average \cite{Fulga2014}. 
We find that the exact self-dual structure of the interaction Hamiltonian protects the gapless Majorana mode up to extremely strong attractive interactions.
A gapped phase does appear for repulsive interactions, but first the Kitaev edge enters a gapless phase with an unusually large central charge $c=3/2$. 

\emph{Interacting Kitaev edge} --- 
The end points of each nanowire (labeled $s=1,2,\ldots$) in the array of 
Fig.~\ref{fig:kitaevedge} form a 1D lattice of self-conjugate Majorana 
operators ($\gamma_s^\pd=\gamma_s^\dag$), governed by the Hamiltonian
\begin{equation}\label{eq:Hfull}
 H = -i  \sum_{s} \alpha_s\gamma_s\gamma_{s+1} - \sum_{s} \kappa_s \gamma_s\gamma_{s+1}\gamma_{s+2}\gamma_{s+3}.
\end{equation}
The $\alpha_s$ terms describe hopping along the edge and the $\kappa_s$ terms describe interactions. Only for odd $s$ do the four-Majorana terms 
have the interpretation of density-density (Hubbard-type) interactions \cite{Chiu2015}. The presence of the even $s=2,4,\dots$ interaction term 
is dicated by the translation symmetry of the mesoscopic array. The
fact that all four-Majorana terms appear on equal footing determines that
the edge Hamiltonian is exactly self-dual. 

We perform a Jordan-Wigner transformation to an equivalent spin-$\frac{1}{2}$ representation, by writing the Majorana operators in terms of Pauli matrices:
\begin{equation}
\gamma_{2k}=\sigma_k^y\, \prod_{j=1}^{k-1}\sigma_j^z,\;\;
\gamma_{2k-1}=\sigma_k^x\, \prod_{j=1}^{k-1}\sigma_j^z.\label{spinMajorana}
\end{equation}
This transformation splits the coupling parameters into even, 
$\alpha_s^{\rm e}\equiv\alpha_{2s}$, $\kappa_s^{\rm e}\equiv\kappa_{2s}$, and odd, $\alpha_s^{\rm o}\equiv\alpha_{2s+1}$, $\kappa_s^{\rm o}\equiv\kappa_{2s+1}$ sets.
Statistical translation invariance dictates that the even and odd sets are indistinguishable in a clean system and have the same probability distribution in a disordered ensemble.

The Hamiltonian \eqref{eq:Hfull} of the interacting Kitaev edge transforms into
\begin{equation}\label{eq:Hspin}
\begin{split}
 H=& -\sum_s \alpha_s^{\rm e} \,\sigma_s^x 
     -\sum_s \alpha_s^{\rm o} \, \sigma_s^z\sigma_{s+1}^z  \\
 & + \sum_s \kappa_s^{\rm e} \,\sigma_s^x\sigma_{s+1}^x
   + \sum_s \kappa_s^{\rm o} \,\sigma_s^z\sigma_{s+2}^z.
\end{split}
\end{equation}
This spin model is the self-dual anisotropic next-nearest-neighbor Ising (ANNNI) model. The standard (\(\kappa_s^{\rm e}=0\)) ANNNI model \cite{Bak1982, Selke1988} has generically a gapped spectrum, except for a critical line with \(c=0.5\) and a floating phase with \(c=1\). 
(See Ref.\ \onlinecite{Hassler2012, Thomale2013} for a recent study in the context of Majorana zero-modes.)
The special feature of the Kitaev edge that protects the gapless phase is the equivalence of the even and odd coupling terms \cite{Fulga2014,Diez2014a}.
The ANNNI Hamiltonian \eqref{eq:Hspin} satisfies a corresponding self-duality relation \cite{Cobanera2010}, which in the context of a spin-$\frac{1}{2}$ chain would require an artificial fine-tuning of parameters.
Here the self-duality is inherited from the realization of the Kitaev edge Hamiltonian \eqref{eq:Hfull}, where it expresses the natural requirement that the translation of the Majorana operators by one site, $\gamma_s\mapsto\gamma_{s+1}$, should describe the same physical system.
Self-duality pins the Kitaev edge at a gapless critical point between two gapped phases, protecting it from localization by disorder or finite but potentially extremely large interactions.

\emph{Phase diagram of the clean edge} --- 
In the absence of disorder we may set $\kappa_s^{\rm e} = \kappa_s^{\rm o} \equiv \kappa$ and $\alpha_s^{\rm e} = \alpha_s^{\rm o} \equiv \alpha\ (>0$ for definiteness \footnote{The choice $\alpha > 0$ can be made without loss of generality, since its sign can be changed by the gauge transformation $\gamma_s \to (-1)^s\gamma_s$ in Majorana language, or $\sigma_s^x\to-\sigma_s^x$ and $\sigma_s^z \to (-1)^s\sigma_s^z$ in spin language.}).
Then the edge is controlled by only one dimensionless interaction-strength parameter $\zeta=\kappa / \alpha$. 

\begin{figure}[tb]
 \includegraphics[width=0.9\columnwidth]{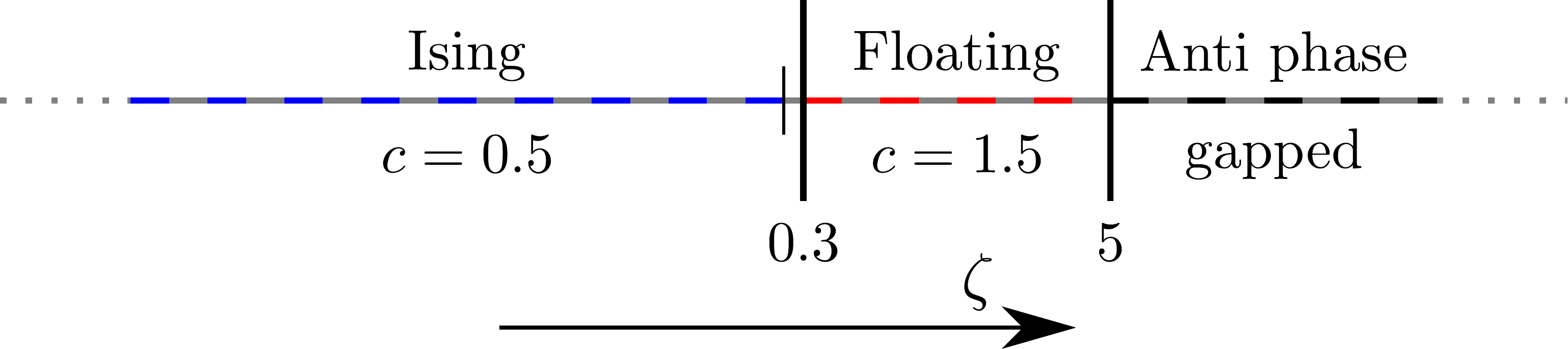}
 \caption{Quantum phase diagram of the interacting Kitaev edge as a function of the relative strength $\zeta$ of interactions and hopping along the edge. The central charges associated with the gapless phases are indicated.
\label{fig:pd}}
\end{figure}

\begin{figure}[tb]
\includegraphics[width=0.8\columnwidth]{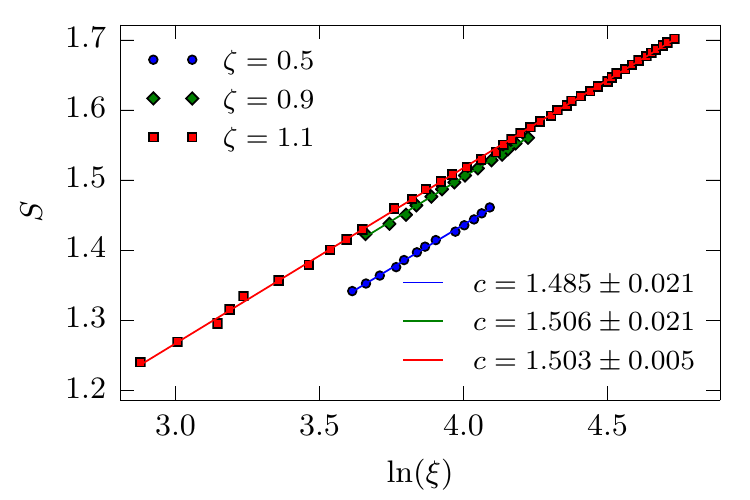}
\caption{Correlation length dependence of the entanglement entropy for different control parameters $\zeta$ in the ``floating'' phase. The straight line fits to $S = (c/6) \ln \xi + {\rm const.}$ indicate a central charge of $c=3/2$. Bond dimensions used are in the range \mbox{$18 \le D \le 102$}. All states were converged up to an effective energy gradient norm $\le 10^{-8}$. 
\label{fig:cc15}}
\end{figure}

We examine the phase diagram of the clean Kitaev edge using 
the \emph{evoMPS} implementation \cite{Haegeman2013, Milsted2013, Milsted2013a} of the MPS time-dependent variational principle and conjugate gradient solver in the thermodynamic limit.
To determine the central charge of the gapless (critical) phases we divide the infinite chain into half-chains $A$ and $B$ and calculate the entanglement entropy \mbox{\(S = - {\rm Tr}\, \rho_A \ln \rho_A\)} from the reduced density matrix $\rho_A$.
The scaling \mbox{$S = (c/6) \ln \xi + \text{const.}$} of $S$ with the correlation length $\xi$ of the slowest decaying correlation function provides an estimate of the central charge \cite{Calabrese2004, tagliacozzo2008}.
A saturation of $S$ with $\xi$ on the other hand provides strong evidence for a gapped phase.
In this way we obtain the three phases indicated in Fig.\ \ref{fig:pd}. 
Representative numerical data for the floating phase are shown in Fig.\, \ref{fig:cc15}.

The non-interacting point $\zeta=0$ corresponds to the critical phase of the Ising model, with central charge $c=1/2$. We find that this phase persists up to extremely large attractive interactions. Our numerics extends up to \(\zeta=-100\), consistent with a very recent independent report \cite{Rahmani2015} of a phase transition into a gapped phase for $\zeta\approx -250$ (see also \cite{supplemental}).

\begin{figure}[b]
 \includegraphics[width=\columnwidth]{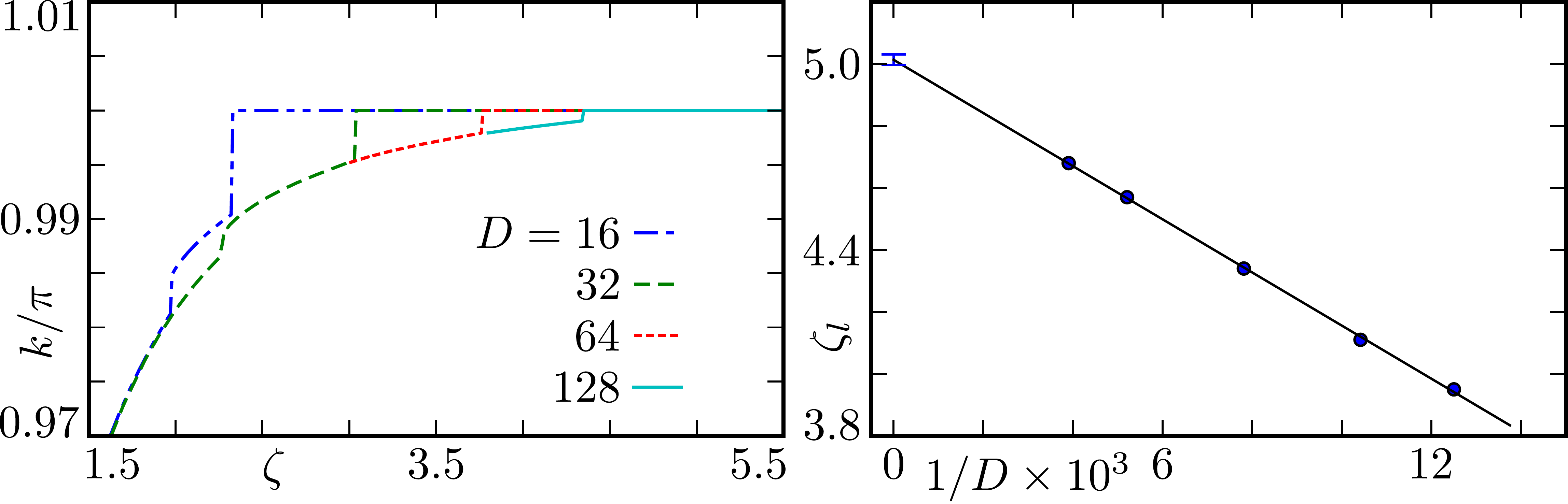}
 \caption{Left panel: Locking of the wavevector corresponding to $\sigma^x$ modulations as a function of the control parameter $\zeta$ for different bond dimensions $D$. Right panel: Extrapolation of the locking point, $\zeta_l$, to infinite bond dimension.\label{fig:icdw}}
\end{figure}

For $\zeta>0$ a second-order phase transition takes place at  $\zeta\approx 0.3$, leading to a new gapless phase characterized by incommensurate charge-density waves. 
We examine the dominant wavevector $k$ corresponding to modulations in the connected $\sigma^x$ and $\sigma^z$ spin-spin correlation functions (see Fig.~\ref{fig:icdw}). 
In the Ising phase (\(\zeta\lesssim 0.3\)) these correlation functions have diverging correlation length and no modulations, so $k=0$. At $\zeta\approx 0.3$ the system enters 
a ``floating" phase in which the wavevector varies continuously. 
Unlike the floating phase of the standard ANNNI model, the floating phase 
of the Kitaev edge has has an unusually large central charge of $c=3/2$, 
see Fig.\, \ref{fig:cc15}..

For sufficiently large \(\zeta\gtrsim 5\), the wavevector locks to a commensurate value: $k=\pi/2$ for $\sigma^z$ and $k=\pi$ for $\sigma^x$. A finite discontinuity in the second derivative of the energy accompanies the locking of the wavevector.
We interpret the locking of the wavevector as a transition into a phase characterized by a commensurate charge-density wave, akin to the gapped anti-phase of the standard ANNNI model. 
The value of $\zeta$ where the locking takes place depends significantly on the size of our numerical simulation (more precisely, on the bond dimension \(D\) of the infinite MPS), and therefore the transition point at $\zeta\approx 5$ is a $D \to \infty$ extrapolation, as shown in Fig.~\ref{fig:icdw}.

\emph{Effect of disorder} --- 
We model disorder in the Kitaev edge by shifting the nearest-neighbor hopping terms $\alpha_s^{\rm o}$ and $\alpha_s^{\rm e}$ by a random amount in the range $[-\delta, \delta]$, drawn independently and uniformly for each lattice site. 
Because $\alpha_s^{\rm o}$ and $\alpha_s^{\rm e}$ are statistically equivalent, the translation invariance on long length scales is not broken by disorder. 
The disorder-averaged entanglement entropy ${\cal S}$ of a delocalized 1D system with open boundaries (divided into segments of length $x$ and $L-x$) is given by \cite{Refael2004,Binosi2007,Refael2009}
\begin{equation}\label{eq:random_entropy}
 {\cal S}(x) = \tfrac{1}{6}\widetilde{c}\ln \left[ (2L/\pi) \sin \left( \pi x/L\right) \right] + {\rm constant}.
\end{equation}
This formula is an adaptation of the well-known clean case \cite{Holzhey1994,Calabrese2004}, with $\widetilde{c}$ an effective central charge instead of the usual central charge $c$ associated with translation-invariant systems.

\begin{figure}[tb]
\includegraphics[width=0.9\columnwidth]{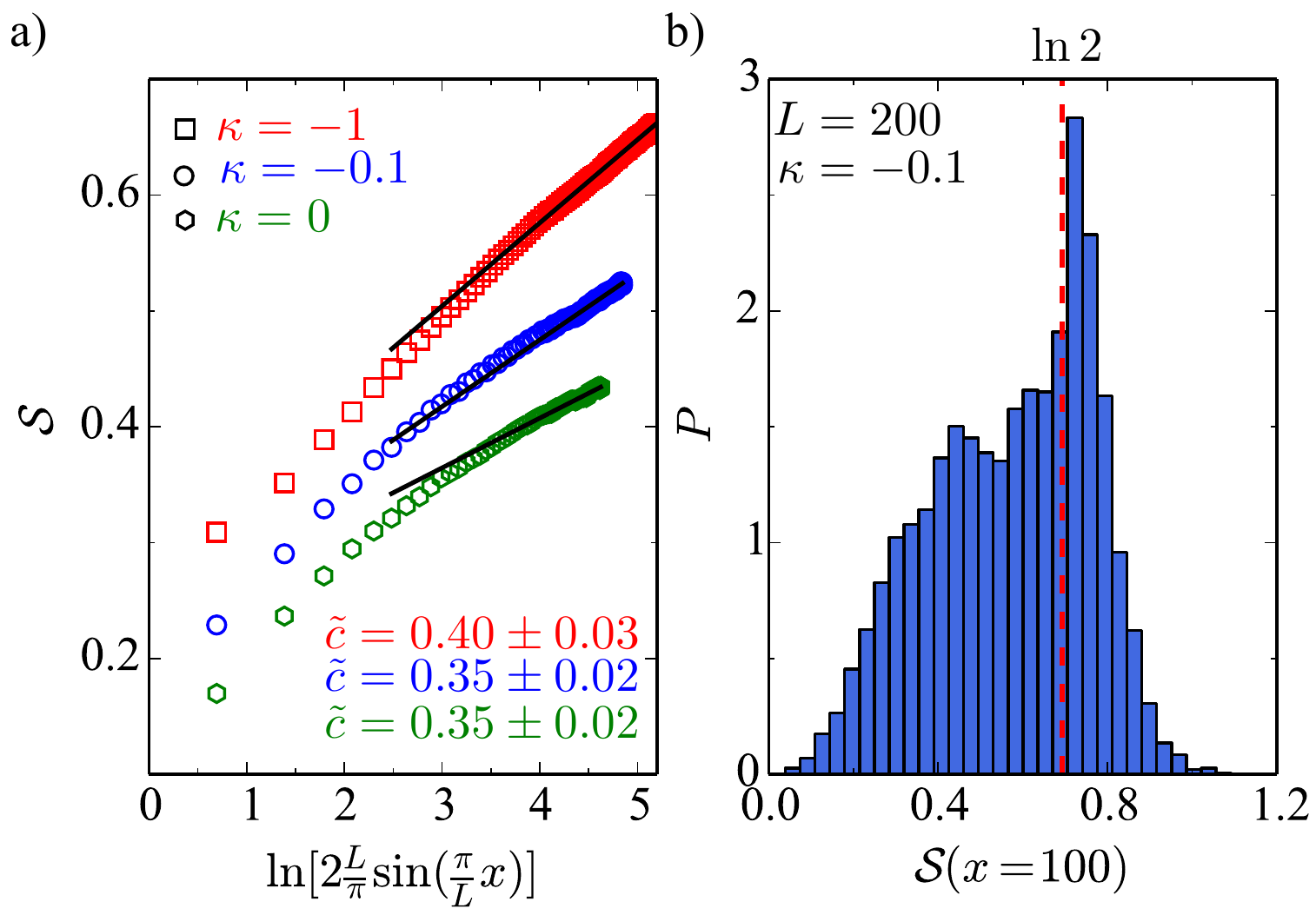}
\includegraphics[width=0.9\columnwidth]{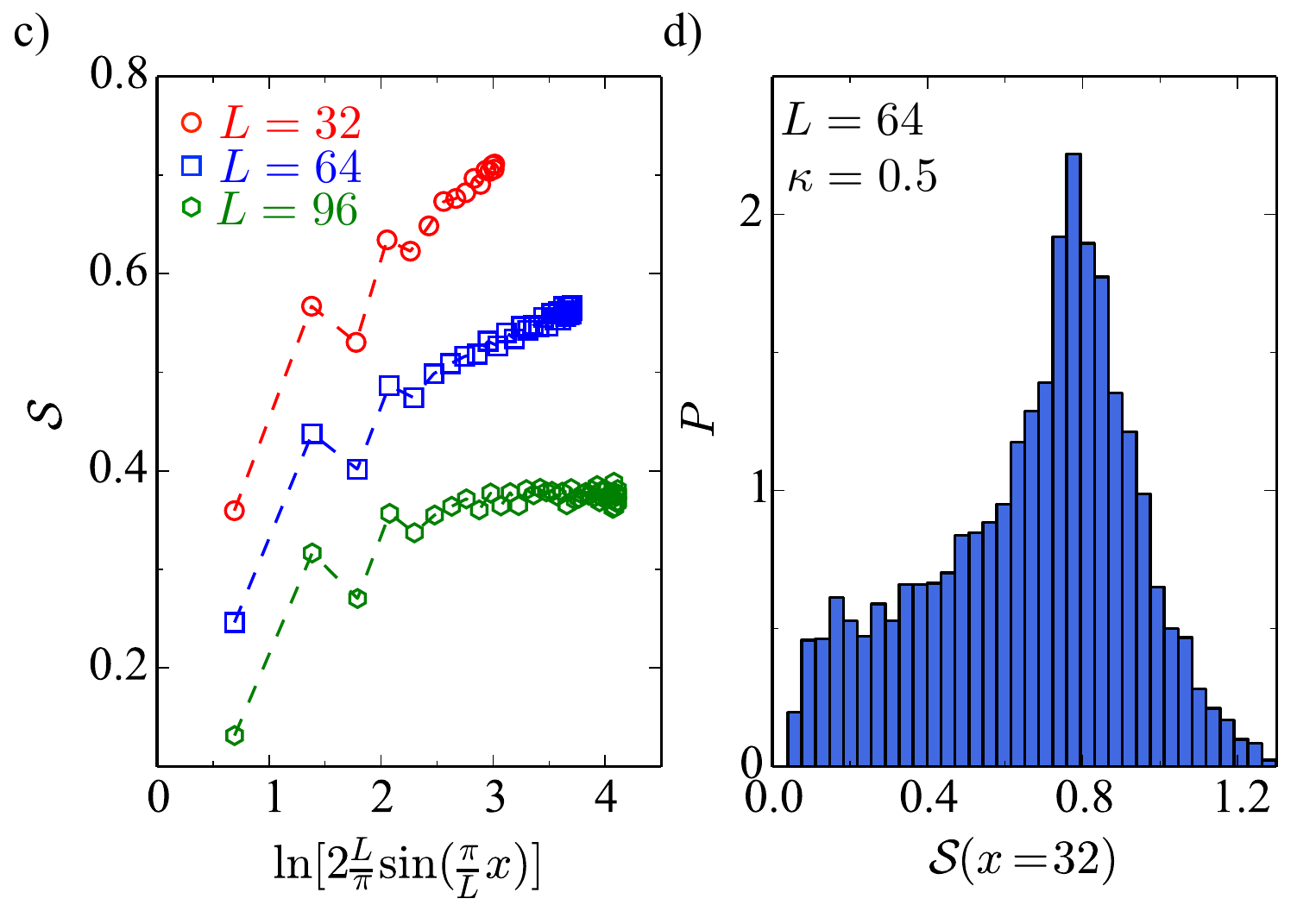}
\caption{
Top panels: Effect of disorder on the Ising phase at $\alpha=1$ with $\delta=0.5$.
(a) Scaling of the average entanglement entropy for attractive interactions $\kappa = 0, -0.1, -1$ at $L=100, 200, 300$ respectively, according to
Eq.~\eqref{eq:random_entropy}. 
The effective central charge is obtained from the slope of linear fits (solid lines) in the limit $x\to L/2$ as $\widetilde{c}/6$. 
Data has been shifted vertically for clarity.
 (b)
Normalized probability distribution of the half-chain entanglement entropy for $\kappa = -0.1$, 
showing a developing narrow singlet peak at $\ln 2$. Bottom panels: Effect of disorder on the floating phase
at $\alpha=1$, $\delta=0.5$, and repulsive interactions $\kappa=0.5$. 
(c) The average entanglement entropy for $L = 32, 64, 96$ saturates for large values of the scaling function. Lines are guides to the eye and the data has been shifted vertically for clarity. 
(d) Normalized probability distribution of the half-chain entanglement entropy for $L = 64$, showing evidence for states with entanglement entropy ${\cal S} \to 0$.
Up to $\sim8000$ disorder realizations were employed for each simulation.
\label{fig:disorder}}
\end{figure}

Previous work \cite{Refael2004,Refael2009} has shown that the addition of disorder to a critical Ising phase ($c = 1/2$) drives the spin system to a gapless phase of random spin-singlets, each contributing $\ln 2$ to the entanglement entropy. 
Characteristic signatures of this phase are a) an effective central charge $\widetilde{c} = \frac{1}{2}\ln 2 \approx 0.347$, and b) the appearance of a peak at $\ln 2$ in the probability distribution of ${\cal S}$ due to the singlet contribution.
In Fig.~\ref{fig:disorder} we search for these signatures, both in the $c=1/2$ Ising phase and in the $c=3/2$ floating phase, using an MPS implementation of the DMRG method~\cite{White1992,Kjall2013}.

For the Kitaev edge at $\zeta = 0$ we find that the effective central charge converges quickly to $\widetilde{c}=\frac{1}{2}\ln 2$, as expected (green data points in Fig.~\ref{fig:disorder}a). For small values of $\zeta<0$, we also observe the same behavior (blue data points in Fig.~\ref{fig:disorder}a) very clearly.
The fact that the Kitaev edge remains delocalized for small $\zeta<0$, or equivalently, that the self-dual ANNNI model remains in the random spin-singlet phase, is further confirmed by the narrow peak developing at $\ln 2$ in the half-chain entropy distribution, see Fig.~\ref{fig:disorder}b. 
Finite-size corrections become more and more significant with increasing $-\zeta$, making it difficult to reach a good convergence for $\zeta=-1$. The red data points in Fig.~\ref{fig:disorder}a give a value of $\widetilde{c}$ which we believe has not yet fully converged for $L=300$, but still seems consitent with a delocalized edge.

Simulations in the floating phase \mbox{(\(0.3<\zeta<5.0\))} are computationally more expensive since the entanglement entropy, and therefore the required MPS bond dimension, is larger in this phase (keeping $L$ fixed) due to the unusually large central charge \(c=1.5\). 
Moreover, in order to avoid  incommensurate spin-spin correlations, it becomes necessary to select values of $\kappa$ which give a family of system sizes $L$ commensurate with the ground-state wave vector.
Figure \ref{fig:disorder}c shows that the disorder-averaged entanglement entropy saturates as the middle of the chain is approached for the largest system sizes available, indicating a gapped phase, albeit with a large value of $\mathcal{S}$. 
The probability distribution of Fig.~\ref{fig:disorder}d also shows a finite weight of the distribution for vanishingly small values of the average entropy, which is another indication of a gapped phase. We conclude that the floating phase is localized by disorder, most likely due to the pinning of incommensurate charge-density waves by the random spatial fluctuations of the disorder potential.

\emph{Summary} --- 
We have investigated a strongly-interacting topological insulator stabilized by a symmetry that is broken locally but restored on average. 
Such a statistical topological insulator may be realized at the edge of an anisotropic \textit{p}-wave superconductor.
Of particular interest is the possibility of realizing the interacting Majorana edge mode studied here starting from a two-dimensional array of vortex lines \cite{Chiu2015}, since this could produce larger interaction strengths than proposals involving semiconducting nanowires or atomic chains proximity coupled to superconductors.
Unlike typical one-dimensional models, the effective edge theory of the system remains critical, even for large attractive interaction strength and/or disorder strength.
This behavior can be traced back to the average translation symmetry of the two-dimensional bulk, which imposes an average self-duality on the strongly-interacting Kitaev edge. 
We hope that our work will motivate the search for other strongly interacting topological phases in which average symmetries of the lattice lead to boundaries which remain delocalized in the presence of disorder.

\emph{Acknowledgements} --- 
We thank A. R. Akhmerov, E. Altman, P. Fendley, M. Maksymenko, Y. Nakata, Z. Nussinov, J. Ruhman, N. Shannon, and T. J. Osborne for stimulating discussions. 
This work is part of the DITP consortium, a program of the Netherlands Organisation for Scientific Research (NWO) that is funded by the Dutch Ministry of Education, Culture and Science (OCW).
ICF thanks the European Research Council under the European Union's Seventh Framework Programme (FP7/2007-2013) / ERC Project MUNATOP, the US-Israel Binational Science Foundation, and the Minerva Foundation 
for support.
AM was supported by the ERC Grants QFTCMPS and SIQS and by the cluster of excellence EXC 201 Quantum Engineering and Space-Time Research.
L.S. was supported by an Ali Kaufmann fellowship and the US-Israel Bi-National Science Foundation.

\AtEndEnvironment{thebibliography}{
\bibitem{supplemental} See the Supplemental Material.
}

\bibliography{sike}

\clearpage

\appendix
\section{Supplemental Material}

We provide more information on the clean Kitaev edge in the singular 
limit $\zeta\to\pm\infty$. The dimensionless Hamiltonian reads
\begin{eqnarray}
H=\pm\sum_{s} \gamma_s^\pd\gamma_{s+1}^\pd\gamma_{s+2}^\pd\gamma_{s+3}^\pd.
\end{eqnarray}
The two signs are connected by a gauge transformation (but only if \(\alpha=0\)),
so we focus on the positive sign. Also, for numerical purposes, it is convenient 
to study the Hamiltonian
\begin{eqnarray}
H=\kappa_{\rm e}\sum_{s=2k} \gamma_s^\pd\gamma_{s+1}^\pd\gamma_{s+2}^\pd\gamma_{s+3}^\pd+
\sum_{s=2k-1} \gamma_s^\pd\gamma_{s+1}^\pd\gamma_{s+2}^\pd\gamma_{s+3}^\pd,
\end{eqnarray}
as a function of \(\kappa_{\rm e}\). 

\begin{widetext}

\begin{figure}[htb]
 \includegraphics[width=0.8\textwidth]{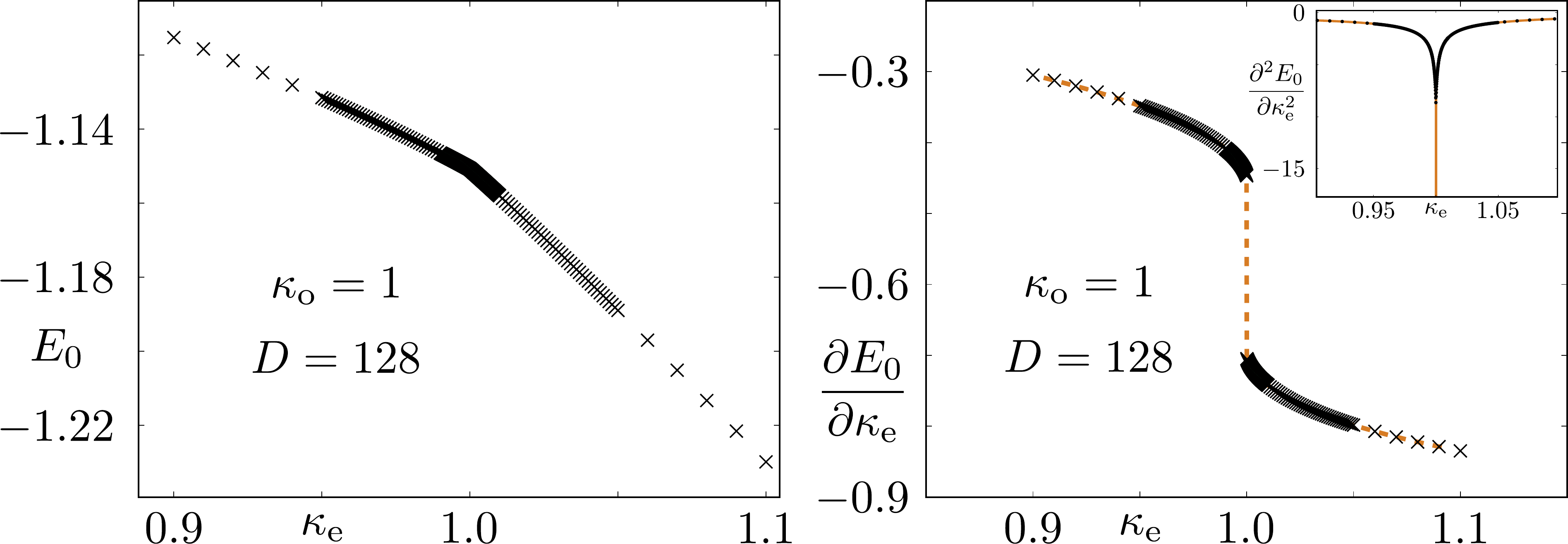}
 \caption{Ground state energy (left panel), as well as its first and second 
derivatives (right panel, inset) as a function of $\kappa_{e}$. We have used 
a constant $\kappa_{\rm o}=1$ and bond dimension $D=128$ throughout. The first 
derivative is discontinuous across the phase transition (dashed line).
\label{fig:firstorder}}
\end{figure}
\end{widetext}

\begin{figure}[htb]
 \centering
  \includegraphics[width=0.9\columnwidth]{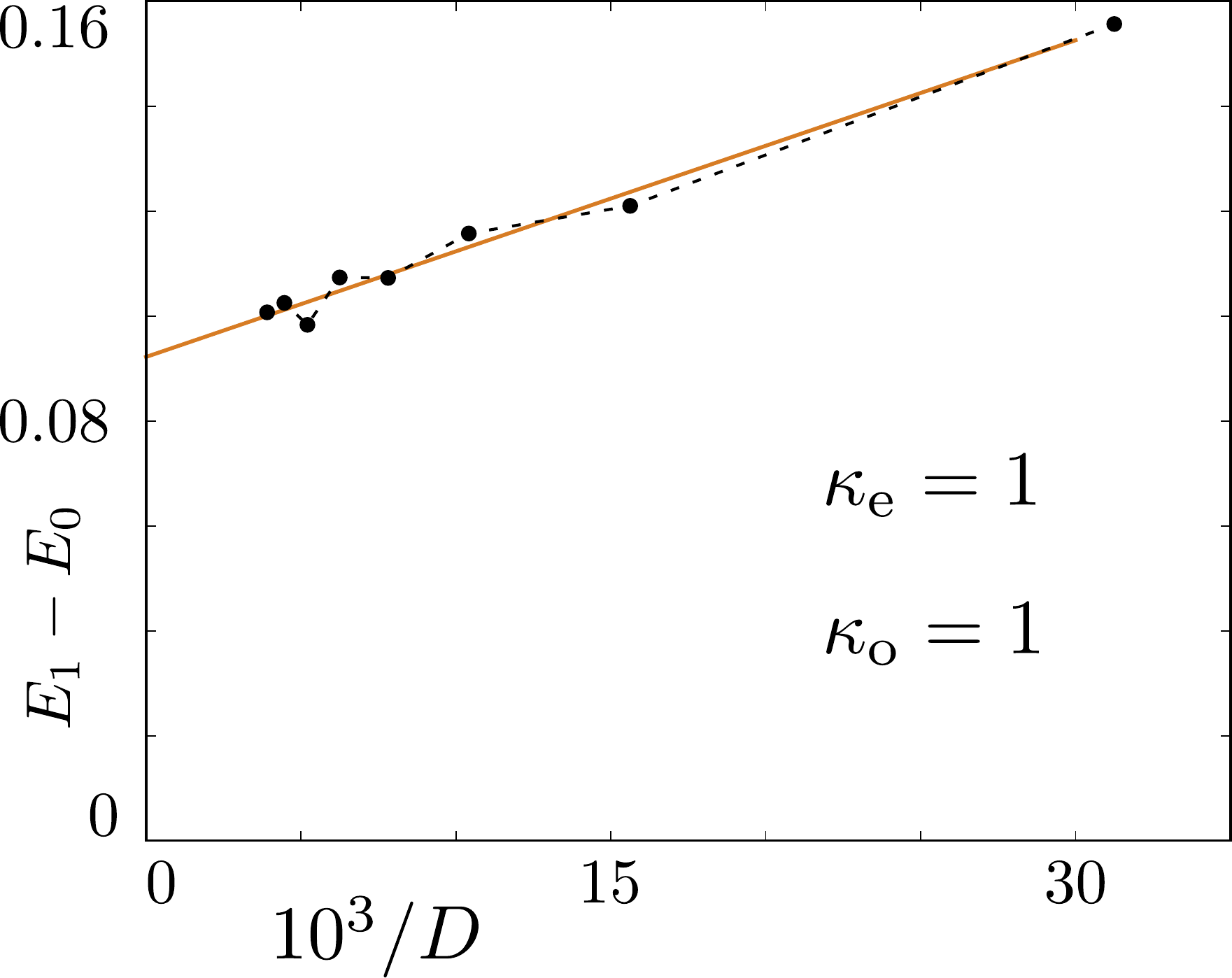}
  \caption{Energy difference between the ground state and the first excited level as a function of inverse bond dimension, $1/D$, at the self-dual point $\kappa_{\rm e}=\kappa_{\rm o}=1$. The system gap scales to a value $E_1-E_0\simeq 0.092$ as the bond dimension is extrapolated to infinity, indicating that the self-dual point is gapped in the thermodynamic limit.\label{fig:gapscaling}}
\end{figure}

Figure~\ref{fig:firstorder} shows the 
behavior of the ground state energy, as well as its first and second derivatives 
as $\kappa_{\rm e}$ is varied through the self-dual point, keeping $\kappa_{\rm o}=1$. 
The first derivative is discontinuous at the phase transition.
To further confirm that the transition at the self-dual point \(\kappa_e=1.0\) 
is of first order, we have performed scaling of the energy gap as a function of 
bond dimension. The results of Fig.~\ref{fig:gapscaling} show that the gap remains 
finite in the thermodynamic limit, in agreement with entanglement entropy results 
(see main text for discussion). Note that the same behavior was reported in the 
recent work of Ref.~\cite{Rahmani2015}.

Since the Kitaev edge is gapped in the limit $\zeta\to\infty$, it is natural
to ask whether this phase of infinitely strong interactions is adiabatically 
connected to the gapped antiphase that we find for \(\zeta>5\). In order
to address this question, we investigated the Kitaev edge for \(\zeta^{-1}\geq 0\).
We conclude that the antiphase does indeed persist all the way to  $\zeta\to\infty$,
and that there is no transition in the singular limit. This situation
should be contrasted with what happens for \(\zeta\to -\infty\), as discussed
in the main text.

\end{document}